\documentclass[10pt]{article}
\usepackage[OE]{express}

\begin{document}
\title{Coherent dynamics of a qubit-oscillator system in a noisy environment}

\author{Wei Wu,\authormark{1*} and Jun-Qing Cheng\authormark{2}}

\address{\authormark{1}Beijing Computational Science Research Center, Beijing 100193, People's Republic of China\\
\authormark{2}Zhejiang Institute of Modern Physics and Physics Department, Zhejiang University, Hangzhou 310027, People's Republic of China}

\email{\authormark{*}weiwu@csrc.ac.cn}

\begin{abstract}
We investigate the non-Markovian dynamics of a qubit-oscillator system embedded in a noisy environment by employing the hierarchical equations of motion approach. It is found that the decoherence rate of the whole qubit-oscillator-bath system can be significantly suppressed by enhancing the coupling strength between the qubit and the harmonic oscillator. Moreover, we find that the non-Markovian memory character of the bath is able to facilitate a robust quantum coherent dynamics in this qubit-oscillator-bath system. Our findings may be used to engineer some tunable coherent manipulations in mesoscopic quantum circuits.
\end{abstract}

\ocis{(030.1640) Coherence; (270.5585) Quantum information and processing.}


\section{Introduction}\label{sec:sec1}

The rapid development of nanotechnology has opened the possibility to realize some quantum optics and quantum information tasks at an atomic scale in experiments. Many current experiments are based on some quantum resources~\cite{1,2}, such as the quantum entanglement~\cite{3}, the quantum correlation~\cite{4} and the quantum coherence~\cite{5,6}. Unfortunately, such quantum resources are quite fragile in the presence of a surrounding environment which destroys the coherence and induces decoherence~\cite{7,8}. The dissipation-induced decoherence in a quantum microscopic system can be effectively modeled by the famous spin-boson model, which describes the interaction between a two-level system and a dissipative bosonic bath~\cite{7,8}. The spin-boson model has attracted considerable attention in past decades because it provides a universal model for numerous physical and chemical processes~\cite{9,10,11}. Due to the fact that the coupling with the surrounding bath is unavoidable, how to fight against the decoherence becomes a very important research field. Various strategies have been proposed to reduce the decoherence in a quantum dissipative system, for example, applying a train of decoupling pulses~\cite{12,13,14,15}, introducing non-linear coupling between the qubit system and its surrounding bath~\cite{16,17,18,19,20,21}, and adding an auxiliary degree of freedom to change the effective bath density spectral function which is responsible for the decoherence behavior~\cite{22,23}. Each scheme has its own regimes of validity depending on the qubit-bath coupling strength, temperature of the bath, as well as the bath density spectral function.

In this paper, motivated by Refs.~\cite{22,23}, we try to achieve a tunable coherent dynamics of a biased qubit-bath model by introducing an auxiliary single-mode harmonic oscillator which acts as a controllable degree of freedom. A similar qubit-oscillator-bath system is also studied in Ref.~\cite{24}, it maybe connect to some photosynthetic pigment-protein complexes. Although the particular configuration of the qubit-oscillator-bath system looks like some non-linear-bath systems proposed in Refs.~\cite{16,18,19,20,25}, it is necessary to point out that the system discussed in this paper is completely different from that of Refs.~\cite{16,18,19,20,25} in which the qubit is designed to be not directly coupled to the multi-mode bosonic environment (the dissipation also enters the qubit dynamics via the intermediate single-mode harmonic oscillator), while, in our study, the qubit is directly embedded in the multi-mode bath and the auxiliary single-mode harmonic oscillator only interacts with the qubit. We also want to emphasize that our study is a nontrivial extension of Refs.~\cite{22,23} in the following two aspects: first, we consider the nonzero bias effect and use a different bath density spectral function. Second, we adopt the hierarchical equations of motion (HEOM) approach~\cite{17,26,27} rather than a perturbative theory to investigate the coherent dynamics of the qubit-oscillator-bath system. The HEOM is a set of time-local differential equations for the reduced density matrix of the quantum subsystem, which was originally proposed by Tanimura and his co-workers~\cite{26}. This numerical treatment includes all the orders of the system-bath interactions and is beyond the usual Markovian approximation, the rotating-wave approximation, and the perturbative approximation. Although there exists a spectrum of numerical techniques for dealing with non-Markovian dynamics in quantum open systems, including the quasiadiabatic path integral method~\cite{28}, the real-time path integral Monte Carlo technique~\cite{29}, the multilayer multiconfiguration time-dependent Hartree theory~\cite{30}, the time-dependent numerical and density matrix renormalization-group approaches~\cite{31}, as well as the Dirac-Frenkel time-dependent variational principle~\cite{32}, HEOM is a well-accepted and highly-efficient numerical formulation. In this sense, compared with the perturbative theory used in Refs.~\cite{22,23}, the conclusion obtained by the HEOM method should be more convincing.

During the past years, there has been an increasing interest to study the memory effect of the bath in quantum dissipative dynamics; this memory effect, which is also called as the non-Markovianity, is a very important characteristic of a quantum open system~\cite{10,33}. How to rigorously define and measure the non-Markovianity has become a hot topic recently. Different theories and physical quantities for detecting the degree of non-Markovian memory effect during a dynamical process are proposed in the previous literatures, for example, the trace distance~\cite{33}, the quantum Fisher information~\cite{34}, the quantum mutual information~\cite{35}, the quantum fidelity~\cite{36}, the quantum channel capacity~\cite{37}, and the $k$ divisibility~\cite{38}. Using the trace distance to characterize the non-Markovian quantum behavior is one of the most popular computable schemes. Trace distance is a metric of the distinguishability between two quantum states~\cite{1,10,33}; the change in the trace distance can be interpreted as a flow of information between the quantum subsystem and the environment~\cite{10,33}. A Markovian process tends to continuously reduce the value of trace distance or, equivalently, the distinguishability between a pair of quantum states, which means that the information flows from the quantum subsystem to its environment. On the other hand, the increase of the distinguishability means a reversed flow of information from the environment back to the quantum subsystem, which is the typical character of a non-Markovian quantum process.

The non-Markovianity is not just a mathematical concept, but has many applications in realistic physical studies~\cite{10,39,40}. In Ref.~\cite{40}, Huelga \emph{et al.} demonstrated that the non-Markovianity is a crucial property that leads to a steady state entanglement, while, a purely Markovian noise would result in the complete destruction of entanglement. A steady state entanglement is a very valuable resource from the view of the experiment, their result implies that the non-Markovianity may be very beneficial to certain quantum optics and quantum information tasks. Considering that fact that the quantum coherence and the quantum entanglement have a close connection, for example, in Ref.~\cite{41}, the authors have proven that any degree of quantum coherence with respect to some reference basis can be converted to entanglement via incoherent operations. Some interesting questions arise naturally here: what is influence of the non-Markovianity on the dynamics of the qubit-oscillator-bath system considered in this paper? Does the non-Markovianity favor the coherent dynamics? In this paper, we try to address this question by connecting the dynamics of the qubit-oscillator-bath system with the rigorous measure of its non-Markovianity in term of trace distance.

This paper is organized as follows. In Sec.\ref{sec:sec2}, we present the qubit-oscillator-bath model which may be feasible in current experiments and briefly outline the general formalism of the HEOM approach. In Sec.\ref{sec:sec3}, we study the non-Markovian dynamics of the qubit-oscillator-bath model and discuss how to the control coherent dynamics by tuning the parameters of the single-mode harmonic oscillator. The main conclusion of this paper is drawn in Sec.\ref{sec:sec4}. Moreover, in Appendix A, we provide some additional details to obtain the steady value of the population difference. In Appendix B, we briefly show how to measure the non-Markovianity in a numerical simulation.

\section{Model and methodology}\label{sec:sec2}

\begin{figure}
\centering
\includegraphics[angle=0,width=12cm]{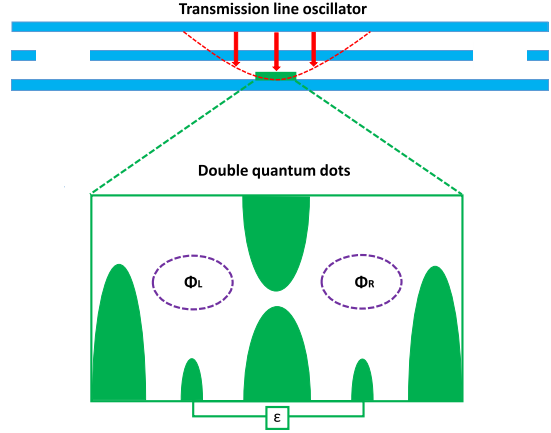}
\caption{\label{fig:fig1} Schematic representation of a possible candidate to realize the qubit-oscillator-bath model in experiments: a double quantum dot couples to an electromagnetic resonator. The biased qubit system can be achieved by the double quantum dot, in which two electrons are confined in a double-well potential created by a two-dimensional electron gas (say a GaAs semiconductor material). By tuning the externally applied voltage and the magnetic field, the two electrons can form one effective spin-qubit. The single-mode harmonic oscillator can be modeled by a transmission line whose frequency is of the order of $\mathrm{GHZ}$ in current experiments. In our study, the qubit-oscillator system is embedded in a noisy bosonic environment, the double quantum dot interacts with the multi-mode bosonic bath via the interdot tunnel gate.}
\end{figure}

We first consider a qubit-oscillator system, in which a biased qubit interacts with a single-mode harmonic oscillator. The Hamiltonian of the qubit-oscillator system is given by
\begin{equation}\label{Eq:Eq1}
\hat{H}_{\mathrm{qo}}=\frac{1}{2}\epsilon\hat{\sigma}_{z}+\frac{1}{2}\Delta\hat{\sigma}_{x}+\omega_{0}\hat{a}^{\dag}\hat{a}+g_{0}\hat{\sigma}_{x}(\hat{a}^{\dag}+\hat{a}),
\end{equation}
where $\hat{\sigma}_{x,z}$ are Pauli matrices to describe the qubit system, $\epsilon$ is the bias field and $\Delta$ denotes the bare tunneling parameter. Operators $\hat{a}^{\dag}$ ($\hat{a}$) is the creation (annihilation) operator of the single-mode harmonic oscillator whose frequency is given by $\omega_{0}$. The coupling strength between the qubit and the single-mode harmonic oscillator is characterize by the parameter $g_{0}$. It is easy to check that $\hat{H}_{\mathrm{qo}}$ is equivalent to the standard biased Rabi model~\cite{42}, by a simple rotation around the $y$ axis.

Such an artificial two-level atom coupled to a harmonic oscillator is one of the most fundamental models for the quantum optics and the quantum information sciences~\cite{43}. Recently, the joint development of artificial atoms and high finesse microwave resonators in nanotechnology has brought the realization of the qubit-oscillator model considered in this paper on a chip. As illustrated in Fig.~\ref{fig:fig1}, the biased qubit can be achieved by making use of a superconducting quantum circuit~\cite{44} or a semiconductor double quantum dot~\cite{45}, while the single-mode harmonic oscillator can be simulated by an electromagnetic transmission line resonator~\cite{46}. Thus, the qubit-oscillator system considered in this paper is not merely of academic interest, but it is a realistic setup in current experiments.

Additionally, we also consider the qubit experiences a dissipative bath which is consists of a set of non-interacting harmonic oscillators. The Hamiltonian of the multi-mode bosonic bath is given by $\hat{H}_{\mathrm{b}}\equiv\sum_{k}\omega_{k}\hat{b}_{k}^{\dag}\hat{b}_{k}$ and the qubit-bath interacting term is described by
\begin{equation}\label{Eq:Eq2}
\hat{H}_{\mathrm{qb}}=\hat{\sigma}_{z}\sum_{k}g_{k}(\hat{b}_{k}^{\dag}+\hat{b}_{k}),
\end{equation}
where $\hat{b}_{k}^{\dag}$ and $\hat{b}_{k}$ are creation and annihilation operators of the $k$th bosonic mode with frequency $\omega_{k}$, respectively. The qubit-bath coupling constant is labeled by the parameter $g_{k}$. Then the total Hamiltonian is written as $\hat{H}=\hat{H}_{\mathrm{qo}}+\hat{H}_{\mathrm{b}}+\hat{H}_{\mathrm{qb}}$. One can find that, the qubit-oscillator-bath system recovers the biased spin-boson model, which has been extensively investigated in many previous articles (see Refs.~\cite{47,48,49} and Appendix A), by dropping the single-mode harmonic oscillator.

Before moving on, we shall introduce our initial-state conditions: we assume the initial state of the whole qubit-oscillator-bath system is a product state: $\hat{\rho}(0)=\hat{\rho}_{\mathrm{q}}(0)\otimes\hat{\rho}_{\mathrm{o}}(0)\otimes\hat{\rho}_{\mathrm{b}}(0)$. Operator $\hat{\rho}_{\mathrm{q}}(0)=|e\rangle\langle e|$ is the initial state of the qubit, where $|e\rangle$ denotes the excited state of the Pauli $z$ operator, i.e., $\hat{\sigma}_{z}|e\rangle=|e\rangle$. Operator $\hat{\rho}_{\mathrm{o}}(0)=|\alpha\rangle\langle\alpha|$ denotes initial state of the single-mode harmonic oscillator, where $|\alpha\rangle\equiv\exp[\alpha(\hat{a}^{\dagger}-\hat{a})]|0_{a}\rangle$ with $\hat{a}|0_{a}\rangle=0$ is the harmonic oscillator's coherent state, the dimensionless parameter $\alpha$ denotes the amount of displacement. The multi-mode bosonic bath is initially prepared in its Fock vacuum state $\hat{\rho}_{\mathrm{b}}(0)=|\mathbf{0}_{\mathrm{b}}\rangle\langle \mathbf{0}_{\mathrm{b}}|$, where $|\mathbf{0}_{\mathrm{b}}\rangle$ is defined by $|\mathbf{0}_{\mathrm{b}}\rangle\equiv\prod_{k}|0_{k}\rangle$ with $\hat{b}_{k}|0_{k}\rangle=0$. In our study, we assume that the multi-mode bosonic bath is always in the equilibrium state, i.e., $\hat{\rho}_{\mathrm{b}}(t)=\hat{\rho}_{\mathrm{b}}(0)=\hat{\rho}_{\mathrm{b}}$.

Generally, it is convenient to encode the frequency dependence of the interaction strengths in the bath density spectral function $J(\omega)\equiv\sum_{k}g_{k}^{2}\delta(\omega_{k}-\omega)$, in this paper, we assume the bath density spectral function $J(\omega)$ has the Lorentz form:
\begin{equation*}
J(\omega)=\frac{1}{2\pi}\frac{\gamma\lambda^{2}}{(\omega-\Delta)^{2}+\lambda^{2}},
\end{equation*}
where $\lambda$ defines the spectral width of the coupling and $\gamma$ can be approximately interpreted as the qubit-bath coupling strength. The reason why we choose the Lorentz spectral function is twofold: firstly, for a Lorentzian bath density spectral function $J(\omega)$, the bath correlation function, $C(t)\equiv\mathrm{tr}_{\mathrm{b}}[\hat{g}_{\mathrm{b}}(t)\hat{g}_{\mathrm{b}}(0)\hat{\rho}_{\mathrm{b}}]$ with $\hat{g}_{\mathrm{b}}(t)\equiv\sum_{k}g_{k}(\hat{b}_{k}^{\dagger}e^{i\omega_{k}t}+\hat{b}_{k}e^{-i\omega_{k}t})$, is given by~\cite{8}
\begin{equation}\label{Eq:Eq3}
\begin{split}
C(t)=&\int d\omega J(\omega)e^{-i\omega t}\\
=&\frac{1}{2}\gamma\lambda e^{-(\lambda+i\Delta)t}.
\end{split}
\end{equation}
This is an Ornstein-Uhlenbeck-type bath correlation function which is the key requirement to perform a HEOM simulation~\cite{17,26,27}. Secondly, the Lorentzian spectral function has a clear boundary between Markovian and non-Markovian regimes~\cite{8}. More specifically speaking, the parameter $\lambda$ is connected to the bath correlation time $\tau_{\mathrm{b}}$ by the relation $\tau_{\mathrm{b}}\simeq \lambda^{-1}$, while the time scale $\tau_{\mathrm{s}}$, on which the state of the system changes, is given by $\tau_{\mathrm{s}}\simeq \gamma^{-1}$. In this sense, the boundary between Markovian regimes and non-Markovian regimes can be approximately specified by the ratio of $\tau_{\mathrm{b}}/\tau_{\mathrm{s}}=\gamma/\lambda$. When $\gamma/\lambda$ is very small, which means the bath correlation time $\tau_{\mathrm{b}}$ is much smaller than the relaxation time of the quantum subsystem $\tau_{\mathrm{s}}$, the decoherence mechanics is Markovian. When $\gamma/\lambda$ is large, the memory effect of the bath should be taken into account and the dynamics in this case is then non-Markovian. In fact, the hierarchical equations are equivalent to the common Markovian Lindblad-type master equation when $\lambda\ll\gamma$, due to the fact that the bath correlation function reduces to the Dirac $\delta$ function, i.e., $C(t-t')\rightarrow \mathrm{const}.\times\delta(t-t')$ in this situation. This feature of the Lorentz spectral function is very helpful for us to build the connect between the non-Markovianity and the coherent dynamics of the qubit-oscillator-bath system. However, we also want to point out that a more rigorous way to distinguish the Markovian or non-Markovian regimes in parameter space should consider not only the bath density spectral function $J(\omega)$ but also the realistic decoherence channel. Thus, we also adopt the scheme proposed by Breuer \emph{et al}.~\cite{33} to measure the degree of non-Markovian in this qubit-oscillator-bath system (see Appendix B for more details). The rigorous result obtained by the trace distance is in good agreement with our above analysis.

In this paper, we adopt the HEOM method to investigate the non-Markovain coherent dynamics. For the Ornstein-Uhlenbeck-type bath correlation function given by Eq.\ref{Eq:Eq3}, the hierarchical equations can be derived from the ordinary Schr$\ddot{\mathrm{o}}$dinger equation $\partial_{t}|\psi(t)\rangle=-i\hat{H}|\psi(t)\rangle$ or the quantum Liouville equation $\partial_{t}\hat{\rho}(t)=-i[\hat{H},\hat{\rho}(t)]$ by making use of the Feynman-Vernon influence functional approach~\cite{26,27} or stochastic perspectives~\cite{17,50}. Following procedures shown in Refs.~\cite{17,50}, one can obtain the hierarchy equations of the reduced qubit-oscillator system (the degrees of freedom of the multi-mode bosonic bath has been partially traced out) as follows:
\begin{equation}\label{Eq:Eq5}
\begin{split}
\frac{d}{dt}\hat{\rho}_{\vec{\ell}}(t)=&(-i\hat{H}_{\mathrm{qo}}^{\times}-\vec{\ell}\cdot\vec{\mu})\hat{\rho}_{\vec{\ell}}(t)+\hat{\Phi}\sum_{p=1}^{2}\hat{\rho}_{\vec{\ell}+\vec{e}_{p}}(t)+\sum_{p=1}^{2}\ell_{p}\hat{\Psi}_{p}\hat{\rho}_{\vec{\ell}-\vec{e}_{p}}(t),
\end{split}
\end{equation}
where vector $\vec{\ell}=(\ell_{1},\ell_{2})$ is a two-dimensional index, $\vec{e}_{1}=(1,0)$, $\vec{e}_{2}=(0,1)$, and $\vec{\mu}=(\lambda+i\Delta,\lambda-i\Delta)$ are two-dimensional vectors, the superoperators $\hat{\Phi}$ and $\hat{\Psi}_{p}$ are defined as follows:
\begin{equation*}
\hat{\Phi}=-i\hat{\sigma}_{z}^{\times},~~~\hat{\Psi}_{p}=\frac{i}{4}\gamma\lambda[(-1)^{p}\hat{\sigma}_{z}^{\circ}-\hat{\sigma}_{z}^{\times}],
\end{equation*}
with $\hat{X}^{\times}\hat{Y}\equiv[\hat{X},\hat{Y}]=\hat{X}\hat{Y}-\hat{Y}\hat{X}$ and $\hat{X}^{\circ}\hat{Y}\equiv\{\hat{X},\hat{Y}\}= \hat{X}\hat{Y}+\hat{Y}\hat{X}$.

The initial-state conditions of the auxiliary matrices are $\hat{\rho}_{\vec{\ell}=\vec{0}}(0)=\hat{\rho}_{\mathrm{q}}(0)\otimes\hat{\rho}_{\mathrm{o}}(0)$ and $\hat{\rho}_{\vec{\ell}\neq \vec{0}}(0)=0$, where $\vec{0}=(0,0)$ is a two-dimensional zero vector. In our numerical simulations, we need to truncate the size of Hilbert space of the single-mode harmonic oscillator as well as number of the hierarchical equations. Unless otherwise stated, we restrict $\hat{H}_{\mathrm{qo}}$ in a $16\times 16$ matrix (8 bosonic modes of the bath are involved). We also set a sufficiently large integer $L$ as the cutoff order of the HEOM, which means all the terms of $\hat{\rho}_{\vec{\ell}}(t)$ with $\ell_{1}+\ell_{2}>L$ are set to be zero, and all the terms of $\hat{\rho}_{\vec{\ell}}(t)$ with $\ell_{1}+\ell_{2}\leq L$ form a closed set of differential equations. All these differential equations can be easily solved
directly by using the traditional Runge-Kutta method. One can keep on increase the hierarchy order $L$ until the final result converges. The reduced density matrix of the qubit can be obtained by simply partially tracing out of the degrees of freedom of the single-mode harmonic oscillator:  $\hat{\rho}_{\mathrm{q}}(t)=\mathrm{tr}_{\mathrm{o}}[\hat{\rho}_{\vec{\ell}=\vec{0}}(t)]$.

\section{Results and discussions}\label{sec:sec3}

\begin{figure}
\centering
\includegraphics[angle=0,width=6cm]{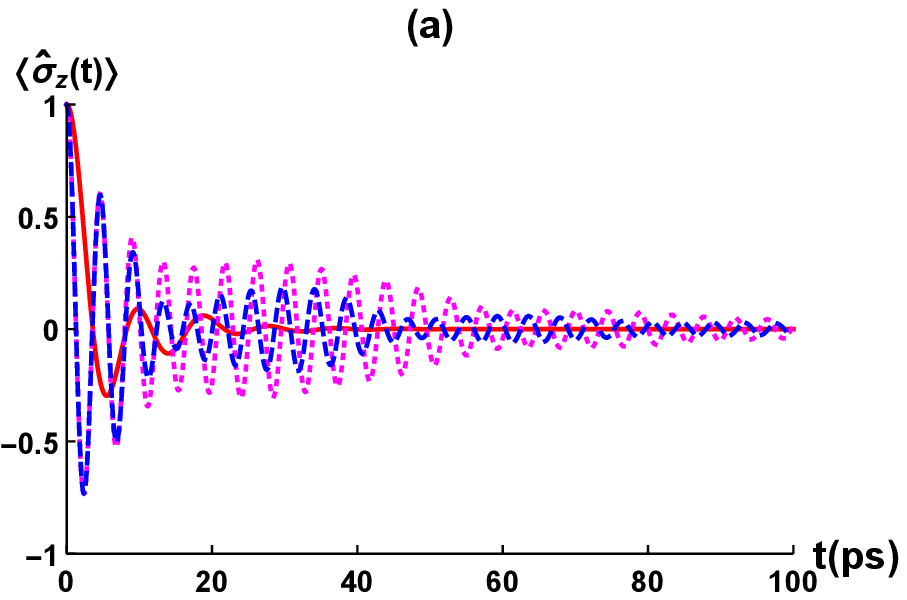}
\includegraphics[angle=0,width=6cm]{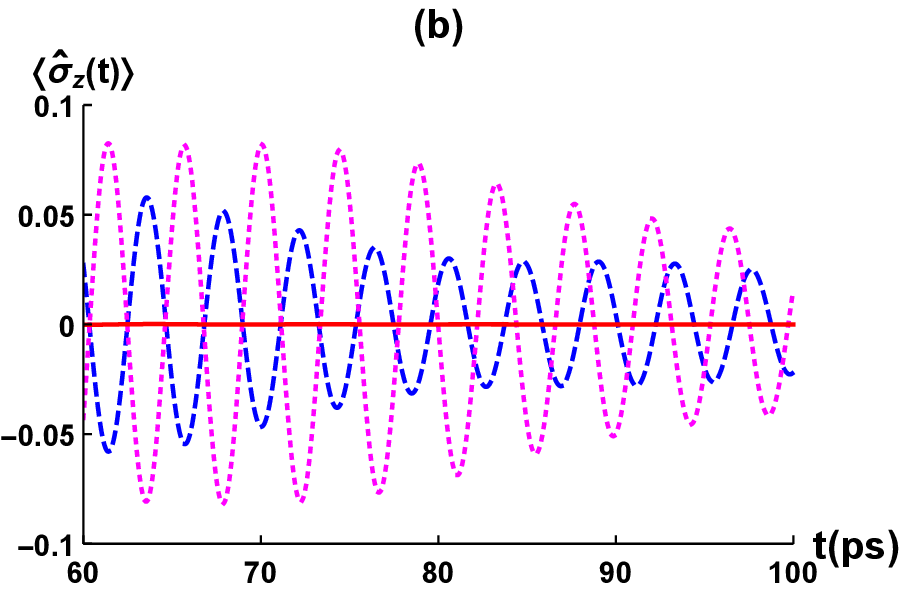}
\includegraphics[angle=0,width=6cm]{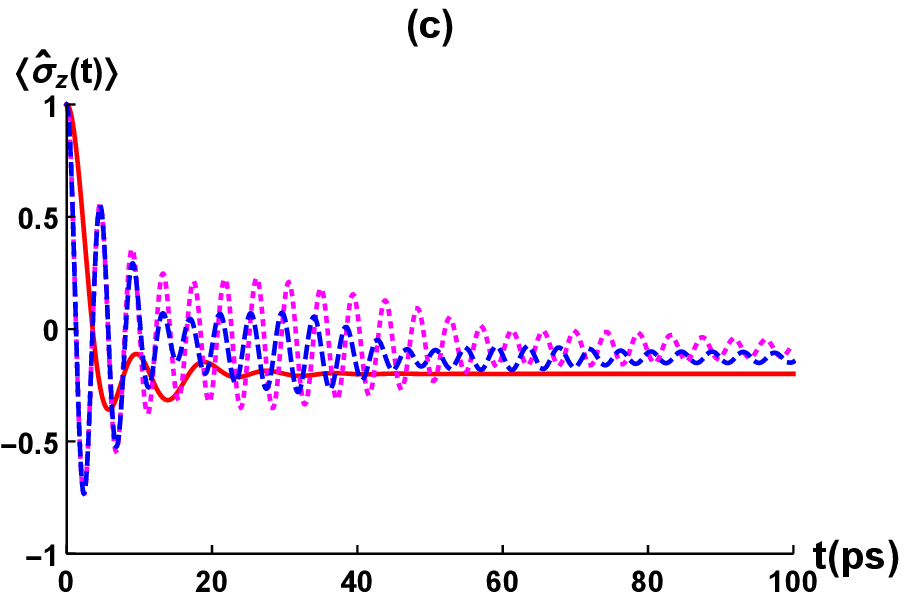}
\includegraphics[angle=0,width=6cm]{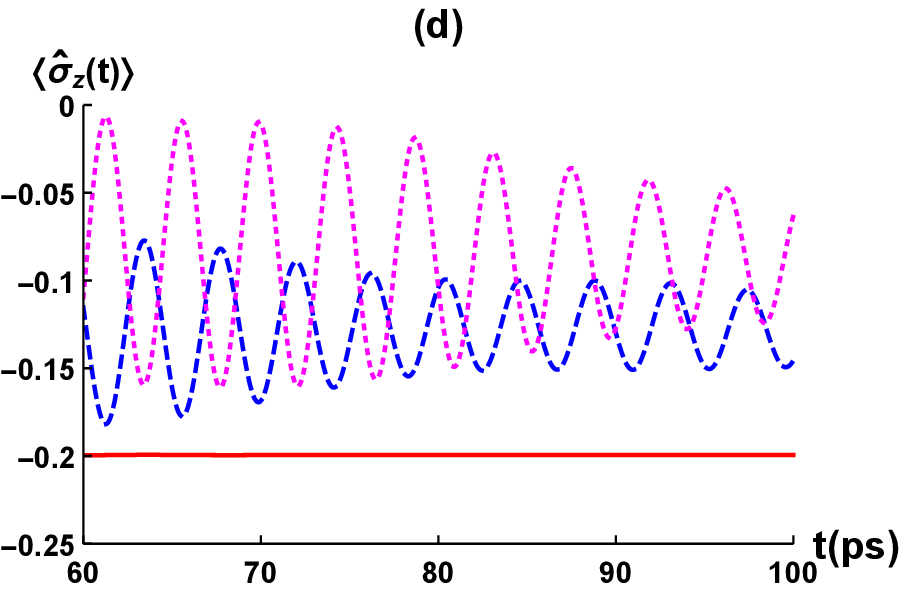}
\caption{\label{fig:fig2} (a) The dynamics of the population difference $\langle\hat{\sigma}_{z}(t)\rangle$ with zero bias $\epsilon=0$ for different parameters: $\zeta=0$ (red solid line), $\zeta=1$ (blue dashed line) and $\zeta=4$ (magenta dotted line). (b) The same with (a), but the range of the time evolution becomes $t\in [60\mathrm{ps},100\mathrm{ps}]$. (c) The dynamics of the population difference $\langle\hat{\sigma}_{z}(t)\rangle$ with nonzero bias $\epsilon=0.1\mathrm{GHZ}$ for different parameters: $\zeta=0$ (red solid line), $\zeta=2$ (blue dashed line) and $\zeta=4$ (magenta dotted line). (d) The same with (c), but the range of the time evolution becomes $t\in [60\mathrm{ps},100\mathrm{ps}]$. Other parameters are chosen as $\alpha=2$, $\gamma=0.5\mathrm{GHZ}$, $\Delta=0.5\mathrm{GHZ}$, $g_{0}=0.02\mathrm{GHZ}$ and $\lambda=0.25\mathrm{GHZ}$.}
\end{figure}
\begin{figure}
\centering
\includegraphics[angle=0,width=6cm]{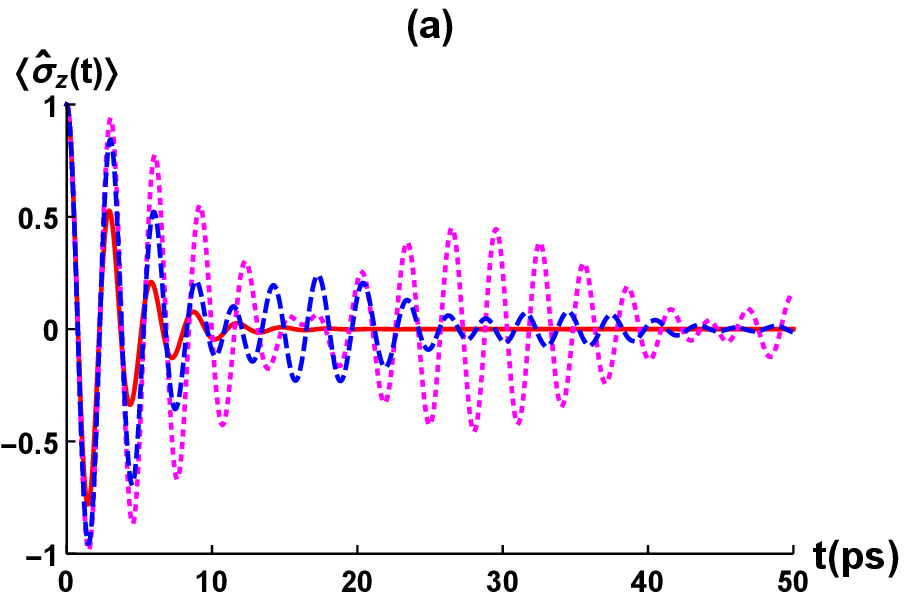}
\includegraphics[angle=0,width=6cm]{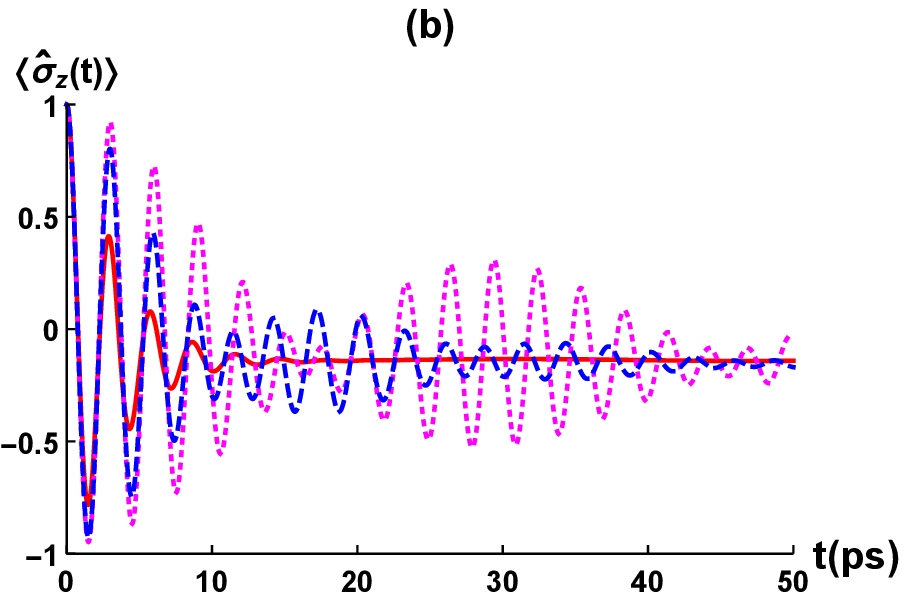}
\includegraphics[angle=0,width=6cm]{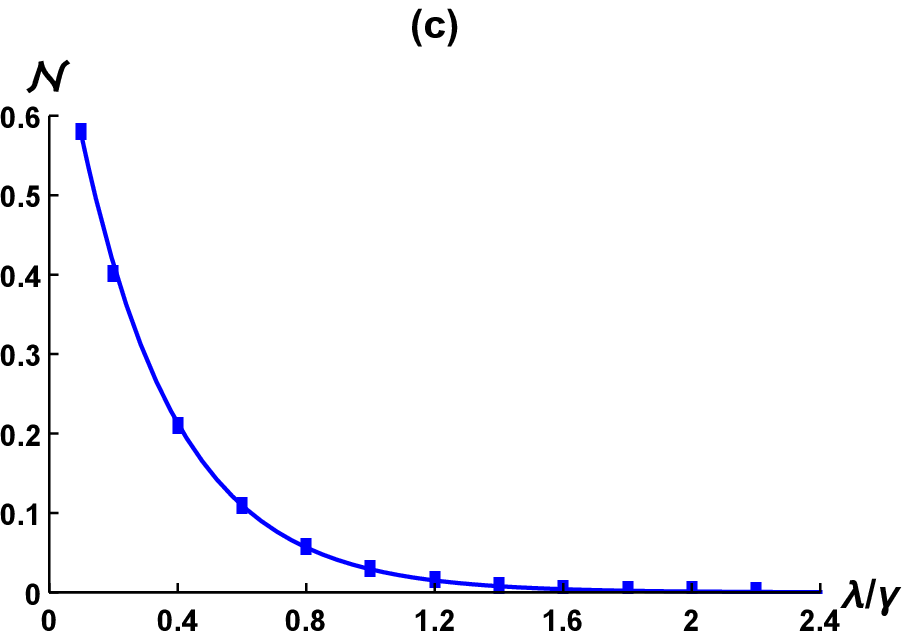}
\includegraphics[angle=0,width=6cm]{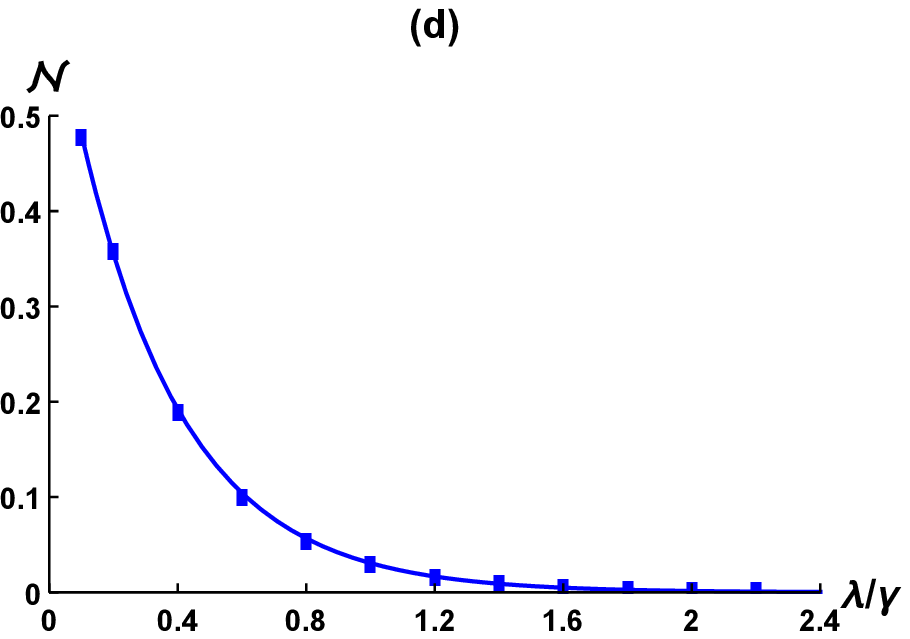}
\caption{\label{fig:fig3} (a) The coherent dynamics of the population difference $\langle\hat{\sigma}_{z}(t)\rangle$ with zero bias $\epsilon=0$ for different spectral widthes: $\lambda=2\gamma$ (red solid line), $\lambda=0.3\gamma$ (blue dashed line) and $\lambda=0.1\gamma$ (magenta dotted line). (b) The same with (a), but $\epsilon=0.3\mathrm{GHZ}$. (c) The non-Markovianity $\mathcal{N}$ as the function of the spectral width with zero bias $\epsilon=0$. (d) The same with (c), but $\epsilon=0.3\mathrm{GHZ}$. Other parameters are chosen as $\gamma=0.5\mathrm{GHZ}$, $\Delta=2\mathrm{GHZ}$, $g_{0}=0.02\mathrm{GHZ}$, $\alpha=2$ and $\zeta=0.1$.}
\end{figure}

In this section, the numerical results of the coherent dynamics of the qubit-oscillator-bath system are presented, we focus on the dynamics of the population difference $\langle\hat{\sigma}_{z}(t)\rangle\equiv \mathrm{tr}_{\mathrm{q}}[\hat{\rho}_{\mathrm{q}}(t)\hat{\sigma}_{z}]$, which is a quantity of interest in experiments.

We start with the zero bias case, i.e., $\epsilon=0$. As one can see from Fig.\ref{fig:fig2}(a), without the single-mode harmonic oscillator ($g_{0}=0$), the population difference $\langle\hat{\sigma}_{z}(t)\rangle$ decays to zero very fast, which indicates the multi-mode bosonic bath destroys the quantum coherence. While, in the present of the single-mode harmonic oscillator, a visible oscillating enhancement is found in the dynamics of the population difference $\langle\hat{\sigma}_{z}(t)\rangle$. Though the oscillation amplitude of $\langle\hat{\sigma}_{z}(t)\rangle$ becomes small as the time evolution, the coherent oscillations persist out to very long times despite the interference of the multi-mode bosonic bath. We also find that such an oscillating enhancement becomes more obvious with the increase of the parameter $\zeta=g_{0}/\omega_{0}$ (one can increase the qubit-oscillator coupling $g_{0}$ with a fixed $\omega_{0}$, or decrease the frequency of the oscillator with a fixed $g_{0}$). This result coincides with previous studies~\cite{22,23}. Our finding implies that the coherent dynamics of the qubit becomes more robust in the dissipative bath by introducing the auxiliary single-mode harmonic oscillator and the coherent dynamics is tunable by adjusting the controllable parameters of the single-mode harmonic oscillator.

For the nonzero bias case, i.e., $\epsilon\neq 0$, the population difference decays to a nonzero steady state value $\langle\hat{\sigma}_{z}(t\rightarrow\infty)\rangle\simeq -\epsilon/\sqrt{\epsilon^{2}+\Delta^{2}}$ with no single-mode harmonic oscillator presents. This numerical result is in agreement with that of Ref.~\cite{47,48}. In the present of the single-mode harmonic oscillator, aparting from the coherent enhancement phenomenon, we observe the steady state value of $\langle\hat{\sigma}_{z}(t\rightarrow\infty)\rangle$ moves away from $-\epsilon/\sqrt{\epsilon^{2}+\Delta^{2}}$ and shifts to a larger value with the increase of $\zeta$. This result can be explained by the frequency renormalization effect: the introducing the auxiliary single-mode harmonic oscillator renormalizes the original bias field's frequency, i.e., $\epsilon\rightarrow\eta\epsilon$, where $\eta$ is the renormalized factor, and results in a larger steady state value (see Appendix A for more details). These result demonstrate that the exist of the single-mode harmonic oscillator can avoid the fast loss of the quantum coherence in a dissipative bath, which is one of basic prerequisites for many tasks in quantum information processing.

Next we explore the influence of the non-Markovian of the multi-mode bath on the coherent dynamics of the qubit. Many previous literatures have shown that the non-Markovianity has significant effects on the decoherence dynamics of a quantum open system~\cite{8,10,11,51}. Does the non-Markovianity favor quantum coherence? Different studies give distinctly different conclusions. For example, in Refs.~\cite{17,52,53}, the authors showed that the non-Markovianity may suppress the decoherence dynamics in the spin-boson model and its extensions. According to their valuations, the Markovain approximation seems result in too much quantum coherence by neglecting the memory effect of the dissipative bath. On the contrary, it has been reported that the non-Markovian effect may favor the quantum coherence dynamics in the three-level system~\cite{54}. Thus, whether the non-Markovianity favor the coherent dynamics or not is still an open issue, different physical systems may lead to different results. From the view of the experiment, it would be very helpful if one can find a quantum system in which the non-Markovianity favors coherent dynamics, because the realistic bath has memory effect, Markovian approximation is just a rough assumption when rigorous results can not achieved.

For the qubit-oscillator-bath system considered in this paper, as we discussed in Sec.\ref{sec:sec2}, the non-Markovianity of the dissipative bath can be simply tuned by changing the ratio of $\lambda/\gamma$: the non-Markovianity becomes weak as the the ratio of $\lambda/\gamma$ increases. This conclusion is carefully checked by using the trace distance as the measure of non-Markovianity, see Appendix B for more details. As shown in Fig.\ref{fig:fig3}, in the non-Markovian regimes, (say $\lambda/\gamma=0.1$ in Fig.\ref{fig:fig3}), the oscillation amplitude is clearly larger than that of the Markovian regime (say $\lambda/\gamma=2$ in Fig.\ref{fig:fig3}). This result demonstrates that the coherent dynamics can be remarkably enhanced by increasing the value of Markovianity and suggests that the non-Markovian character of the bath can be viewed as a valuable quantum resource that favors the coherent dynamics.

\section{Conclusion}\label{sec:sec4}

In our theoretical formalism, the inclusion of the single-mode harmonic oscillator complicates the coherent dynamics considerably, and the ratio of $\zeta=g_{0}/\omega_{0}$ plays a crucial role in this treatment. How to achieve a large value of $\zeta$ is the main difficulty in realizing the coherent control scheme considered in this paper. Thanks to the rapid development in experiments, researchers are able to simulate the Rabi model in the ultra-strong coupling ($g_{0}\geq 0.1\omega_{0}$) and the deep-strong coupling ($g_{0}\sim\omega_{0}$ and $g_{0}>\omega_{0}$) regimes~\cite{add1,add2}. For example, by making use of a superconducting flux qubit and an LC oscillator via Josephson junctions, Yoshihara \emph{et al}. have realized the quantum circuits with the ratio $g_{0}/\omega_{0}$ ranging from $0.72$ and $1.34$~\cite{add2}. As we displayed in Fig.\ref{fig:fig2}(a), when the ratio $g_{0}/\omega_{0}=1$, the effect of the quantum coherence enhancement is very evident.

In summary, we investigate the non-Markovian dynamics of a qubit-oscillator system, built of a biased qubit and a single-mode harmonic oscillator, in a noisy environment by employing HEOM approach. Different from that of Refs.~\cite{16,17,18,19,20,21}, the qubit is directly couples to the bosonic bath and the single-mode harmonic oscillator is introduced as a controllable degree of freedom. We find that the decoherence of the qubit can be reduced by tuning the ratio of $g_{0}/\omega_{0}$ which means the interplay between the combined single-mode harmonic oscillator and the qubit can expand the coherent regime. Moreover, by using the trace distance as the measure of the non-Markovian character of the bath, we find that the non-Markovianity is able to facilitate a robust quantum coherence in this qubit-oscillator-bath system. This result suggests that the non-Markovianity may be utilized as a tool to suppress the decoherence in quantum dissipative systems. In this sense, the non-Markovianity is not just a mathematical concept, but a meaningful quantum resource in quantum manipulations, a similar conclusion also reported in Ref.~\cite{55}. Finally, due to the generality of the qubit-oscillator model, we expect our results to be of interest for a wide range of experimental applications in quantum optics and quantum information science.

\section{Appendix A: The population difference in the long-time limit}\label{sec:sec5}

In this appendix, we briefly show how to approximately obtain the equilibrium steady value of the population difference, i.e., $\langle\hat{\sigma}_{z}(t\rightarrow\infty)\rangle$, by making use of the unitary transformation method which is extensively used in the studies of the Rabi model\cite{56} and the spin-boson model\cite{16,23,47,48}.

First, we apply a unitary transformation to the original Hamiltonian $\hat{H}$ as $\hat{H}'=e^{\hat{S}}\hat{H}e^{-\hat{S}}$, where the generator $\hat{S}$ is given by $\hat{S}=\zeta\hat{\sigma}_{x}(\hat{a}^{\dag}-\hat{a})$ with a undetermined parameter $\zeta$. By using the Baker-Campbell-Hausdorff formula, the transformed Hamiltonian $\hat{H}'$ can be exactly derived as follows
\begin{equation}\label{Eq:Eq6}
\begin{split}
\hat{H}'=&\frac{1}{2}\epsilon\{\cosh[2\zeta(\hat{a}^{\dag}-\hat{a})]\hat{\sigma}_{z}-i\sinh[2\zeta(\hat{a}^{\dag}-\hat{a})]\hat{\sigma}_{y}\}\\
&+\frac{1}{2}\Delta\hat{\sigma}_{x}+\omega_{0}\hat{a}^{\dag}\hat{a}+(g_{0}-\zeta\omega_{0})\hat{\sigma}_{x}(\hat{a}^{\dag}+\hat{a})+\zeta(\zeta\omega_{0}-2 g_{0})\\
&+\sum_{k}\omega_{k}\hat{b}_{k}^{\dag}\hat{b}_{k}+\{\cosh[2\zeta(\hat{a}^{\dag}-\hat{a})]\hat{\sigma}_{z}-i\sinh[2\zeta(\hat{a}^{\dag}-\hat{a})]\hat{\sigma}_{y}\}\sum_{k}g_{k}(\hat{b}_{k}^{\dag}+\hat{b}_{k}).
\end{split}
\end{equation}
If we choose $\zeta=g_{0}/\omega_{0}$~\cite{23,56}, the qubit-oscillator coupling term vanishes, in this case, the transformed Hamiltonian can be rewritten as $\hat{H}'=\hat{H}'_{0}+\hat{H}'_{1}$, where
\begin{equation}\label{Eq:Eq6}
\hat{H}'_{0}=\frac{1}{2}\eta\epsilon\hat{\sigma}_{z}+\frac{1}{2}\Delta\hat{\sigma}_{x}+\sum_{k}\omega_{k}\hat{b}_{k}^{\dag}\hat{b}_{k}+\eta\hat{\sigma}_{z}\sum_{k}g_{k}(\hat{b}_{k}^{\dag}+\hat{b}_{k})+\omega_{0}|\alpha|^{2}-\frac{g_{0}^{2}}{\omega_{0}},
\end{equation}
\begin{equation}\label{Eq:Eq7}
\begin{split}
\hat{H}'_{1}=&\frac{1}{2}\epsilon\{\cosh[2\zeta(\hat{a}^{\dag}-\hat{a})]-\eta\}\hat{\sigma}_{z}+\{\cosh[2\zeta(\hat{a}^{\dag}-\hat{a})]-\eta\}\hat{\sigma}_{z}\sum_{k}g_{k}(\hat{b}_{k}^{\dag}+\hat{b}_{k})\\
&+\omega_{0}\hat{a}^{\dag}\hat{a}-i\epsilon\sinh[2\zeta(\hat{a}^{\dag}-\hat{a})]\hat{\sigma}_{y}-i\sinh[2\zeta(\hat{a}^{\dag}-\hat{a})]\hat{\sigma}_{y}\sum_{k}g_{k}(\hat{b}_{k}^{\dag}+\hat{b}_{k})-\omega_{0}|\alpha|^{2}.
\end{split}
\end{equation}

In this unitary transformation treatment, the parameter $\eta$ is constructed such that $\langle\hat{H}'_{1}\rangle_{\mathrm{o}}=0$~\cite{47,48,57}, where $\langle \hat{H}'_{1}\rangle_{\mathrm{o}}\equiv \mathrm{tr}_{\mathrm{o}}(\hat{\rho}_{\mathrm{o}}\hat{H}'_{1})/\mathrm{tr}_{\mathrm{o}}(\hat{\rho}_{\mathrm{o}})$ denotes the thermodynamic average value with respect to the equilibrium state of the single-mode harmonic oscillator. In an approximate estimation, one can regard $\hat{\rho}_{\mathrm{o}}=\hat{\rho}_{\mathrm{o}}(0)=|\alpha\rangle\langle \alpha|$, then the condition $\langle\hat{H}'_{1}\rangle_{\mathrm{o}}=0$ reduces to $\langle\alpha|\hat{H}'_{1}|\alpha\rangle=0$. It is easy to demonstrate that
\begin{equation*}
\langle \alpha|\sinh[2\zeta(\hat{a}^{\dag}-\hat{a})]|\alpha\rangle=0;~~~\langle\alpha|\hat{a}^{\dag}\hat{a}|\alpha\rangle=|\alpha|^{2}.
\end{equation*}
Then, if we choose the parameter $\eta$ as follows
\begin{equation}\label{Eq:Eq8}
\begin{split}
\eta=&\langle 0_{a}|e^{-\alpha (\hat{a}^{\dag}-\hat{a})}\cosh[2\zeta(\hat{a}^{\dag}-\hat{a})]e^{\alpha (\hat{a}^{\dag}-\hat{a})}|0_{a}\rangle\\
=&\langle 0_{a}|\cosh[2\zeta(\hat{a}^{\dag}-\hat{a})]|0_{a}\rangle\\
=&\exp\Bigg{[}-2\Big{(}\frac{g_{0}}{\omega_{0}}\Big{)}^{2}\Bigg{]},
\end{split}
\end{equation}
one can find that the condition $\langle\alpha|\hat{H}'_{1}|\alpha\rangle=0$ automatically statisfied. In a perturbative treatment, $\hat{H}'_{1}$ can be neglected, then the effective system is given by $\hat{H}'_{\mathrm{eff}}=\hat{H}'_{0}$. It is quite clear to see the effect of the parameter $\eta$ is a renormalization factor of $\epsilon$ which renormalizes $\epsilon$ to a smaller value, because $\eta$ is a real number and smaller than one.

Then, the effective system $\hat{H}'_{\mathrm{eff}}$ reduces to the well-known biased spin-boson model and has been widely studied by a variety of different methods. Due to the fact that the last two terms in Eq.\ref{Eq:Eq6} are constant which have no influences on the dynamical results, one can safely drop them.  Following the detailed exposition in Refs.~\cite{47,48,57}, one can obtain the equilibrium steady value of the population difference of the biased spin-boson model that is given by
\begin{equation}\label{Eq:Eq9}
\begin{split}
\langle\hat{\sigma}_{z}(t\rightarrow\infty)\rangle\simeq&-\frac{\eta\epsilon}{\sqrt{\Delta^{2}+(\eta\epsilon)^{2}}}\\
=&-\cfrac{\epsilon}{\sqrt{\Big{(}\cfrac{\Delta}{\eta}\Big{)}^{2}+\epsilon^{2}}}.
\end{split}
\end{equation}
From the above expression, it is easy to check that a larger value of $\zeta$ makes a larger steady value of the population difference $\langle\hat{\sigma}_{z}(t\rightarrow\infty)\rangle$ which is in agreement with our numerical simulation in Fig.\ref{fig:fig2}.

\section{Appendix B: The measure of non-Markovianity}\label{sec:sec6}

In this appendix, we would like to show how to characterize the non-Markovianity in a quantum open system by making use of the trace distance. According to Ref.~\cite{33}, the most important feature of the non-Markovianity is the emergence of the recoherence or the information backflow which can be described by the rate of the trace distance between two physical initial states. The trace distance of two quantum states $\hat{\rho}_{1}$ and $\hat{\rho}_{2}$ is defined by $D(\hat{\rho}_{1},\hat{\rho}_{2})\equiv\frac{1}{2}\|\hat{\rho}_{1}-\hat{\rho}_{2}\|_{1}$, where $\|\hat{X}\|_{1}\equiv\mathrm{tr}\sqrt{\hat{X}^{\dagger}\hat{X}}$ is the trace norm or the Schatten one-norm of an arbitrary operator $\hat{X}$. For a initial-state pair $\hat{\rho}_{1,2}(0)$ and a given dynamical map $\hat{\Lambda}_{t}$ that generates the time-evolution $\hat{\rho}(t)=\hat{\Lambda}_{t}[\hat{\rho}(0)]$, one can define the rate of change of the trace distance as follows $\varpi[t;\hat{\rho}_{1,2}(0)]\equiv\frac{d}{dt}D[\hat{\rho}_{1}(t),\hat{\rho}_{2}(t)]$. When $\varpi[t;\hat{\rho}_{1,2}(0)]<0$, $\hat{\rho}_{1}(t)$ and $\hat{\rho}_{2}(t)$ become indistinguishable, and this can be understood as the quantum information flows from the quantum subsystem to the environment; when $\varpi[t;\hat{\rho}_{1,2}(0)]>0$, $\hat{\rho}_{1}(t)$ and $\hat{\rho}_{2}(t)$  become distinguishable, and this can be interpreted as the quantum information flows back to the quantum subsystem. Under this spirit, a measure for the non-Markovianity of a quantum process can be defined by
\begin{equation}\label{Eq:Eq10}
\mathcal{N}\equiv\max_{\hat{\rho}_{1,2}(0)}\int_{\varpi>0}dt\varpi[t;\hat{\rho}_{1,2}(0)],
\end{equation}
where the time-integration is extended over all time intervals $t\in[0,+\infty)$ during which $\varpi[t;\hat{\rho}_{1,2}(0)]$ is positive, and the maximum runs over all possible initial-state pairs $\hat{\rho}_{1,2}(0)$.

The definition of non-Markovianity in Eq.\ref{Eq:Eq10} is not suitable for a numerical simulation, thus some slight modifications are needed. First, we only focus on the non-Markovianity accumulated during a finite time interval $t\in[0,t_{c}]$, where $t_{c}$ is the upper bound of the time-integration, because it is impossible to numerically simulate the dynamics of the qubit from zero to $+\infty$. Second, we change the original integrand and its corresponding integrating intervals in Eq.\ref{Eq:Eq10} by a simple algebra that would not change the value of the non-Markovianity, by doing so, one can avoid estimating whether or not $\varpi[t_{i};\hat{\rho}_{1,2}(0)]$ is positive or not at each given time $t=t_{i}$. Finally, a equivalent expression of the non-Markovianity can be written as follows
\begin{equation}\label{Eq:Eq11}
\mathcal{N}\equiv\max_{\hat{\rho}_{1,2}(0)}\frac{1}{2}\int_{0}^{t_{c}}dt\big{\{}|\varpi[t;\hat{\rho}_{1,2}(0)]|+\varpi[t;\hat{\rho}_{1,2}(0)]\big{\}}.
\end{equation}
It is neceaasry to point out that these modifications are also widely adopted in many previous studies~\cite{58,59}.

According to Refs.~\cite{10,25,58,59,60}, the calculation of the non-Markovianity $\mathcal{N}$ can be further simplified by choosing a pair of initial state as two orthogonal states that lie on the boundary of the space of physical states. For the qubit-system case, this orthogonality implies that both of the two initial states must be pure states. Thus, in our numerical simulations, we assume the expressions of these initial-state pairs of the qubit are given by $\hat{\rho}_{\mathrm{q}1}(0)=|\varphi_{\mathrm{q}}(0)\rangle\langle\varphi_{\mathrm{q}}(0)|$ and $\hat{\rho}_{\mathrm{q}2}(0)=|\varphi_{\mathrm{q}}^{\perp}(0)\rangle\langle\varphi_{\mathrm{q}}^{\perp}(0)|$ with
\begin{equation}\label{Eq:Eq12}
|\varphi_{\mathrm{q}}(0)\rangle=\cos\Big{(}\frac{\theta}{2}\Big{)}|e\rangle+e^{i\phi}\sin\Big{(}\frac{\theta}{2}\Big{)}|g\rangle,
\end{equation}
and
\begin{equation}\label{Eq:Eq13}
|\varphi_{\mathrm{q}}^{\perp}(0)\rangle=\sin\Big{(}\frac{\theta}{2}\Big{)}|e\rangle-e^{i\phi}\cos\Big{(}\frac{\theta}{2}\Big{)}|g\rangle,
\end{equation}
where $|e\rangle$ and $|g\rangle$ are the excited and the ground states of Pauli $z$ operator $\hat{\sigma}_{z}$, respectively, the ranges of the two initial-state parameters are given by $\theta\in[0,\pi]$ and $\phi\in[0,2\pi]$. By randomly generating a sufficiently large sample of initial-state parameter combinations $(\theta_{i},\phi_{i})$, one can find the optimal initial-state pair for the non-Markovianity. According to the definition in Eq.\ref{Eq:Eq10}, the measure $\mathcal{N}$ is non-negative, and we have $\mathcal{N}=0$ if and only if the process is Markovian. A nonzero value $\mathcal{N}>0$, implies a non-Markovian process. In our simulation, the upper
bound of the time integration is $t_{c}=50\mathrm{ps}$, and the single-mode harmonic oscillator is approximately regarded as a $4\times 4$ matrix due to the limitation of our computation resources.

\section*{Funding}

This project is supported by the NSFC (Grant No.11704025), the NSAF (Grant No. U1530401) and the China Postdoctoral Science Foundation (Grant No.2017M610753).

\section*{Acknowledgments}

W. Wu wishes to thank Dr. Yi-Nan Fang, Dr. Li-Jing Jin, Dr. Maoxin Liu, Professor Zhiguo Lv and Professor Hai-Qing Lin for many useful discussions.

\section*{Disclosures}

The authors declare that there are no conflicts of interest related to this article.

\end{document}